# Implied volatility formula of European Power Option Pricing

Jingwei Liu[*]   Xing Chen

(School of Mathematics and System Sciences, Beihang University,
LMIB of the Ministry of Education,, Beijing, 100191, P.R China)

**Abstract**：We derive the implied volatility estimation formula in European power call options pricing, where the payoff functions are in the form of $V = \left(S_T^{\alpha} - K\right)^+$ and $V = \left(S_T^{\alpha} - K^{\alpha}\right)^+$ ($\alpha > 0$) respectively. Using quadratic Taylor approximations, We develop the computing formula of implied volatility in European power call option and extend the traditional implied volatility formula of Charles J.Corrado, *et al* (1996) to general power option pricing. And the Monte-Carlo simulations are also given.

## 1. Introduction

In recent decades, financial derivatives pricing attracts much more attention in both economic and statistical fields. For the practical purpose, implied volatility, which estimates the level of financial derivative's risk, is a most important parameter in the Black-Scholes European option pricing model and Merton's European option pricing model [1,2].

As volatility is a measure of uncertainty of the price trend for the future, many works address the problem, and develop different strategies. In 1976, Latane and Rendleman suggested to use implied volatilities in financial markets research[3]. Butler and Schachter (1986) presented an estimator of the Black-Scholes option pricing forluma by Taylor series expansion of the Black-Scholes formula [4]. Chaudhury (1996) proposed another Taylor expansion method to replace the Taylor expansion of Butler and Schachter [5]. Using quadratic Taylor approximations, Corrado and Miller (1996) obtained a close formula of implied volatility estimation[6]. Utilizing the third order Taylor series expansion, Li (2005) developed a new close formula of implied volatility [6]. And, the simulation result of [7] showed that Li's formula is significantly better than the Corrado–Miller formula. However, Li's formula is also more complex than Corrado–Miller formula.

European power option pricing is a hot research field of financial derivative option pricing [8]. In this paper, we derive a new formula to compute European power option implied volatility in the research framework of Corrado and Miller(1996)[6], and give close formula of implied volatility in the power option pricing framework of Liu (2007) [8].

The rest of the paper is organized as follows. In section 2, European power call option pricing formula is introduced. In section 3, the implied volatility estimation formulae are derived. In section 4, the Monte-Carlo simulations are given. The conclusion is given in section 5.

## 2. European power option
### 2.1 Classical European option pricing formula

In the classical risk-neutral market $\left(\Omega, F, \{F_t\}_{t \geq 0}, P\right)$, the price of an asset $S_t$ at time $t$ is supposed to be a geometric Brownian motion,

---

[*] Corresponding author: jwliu@buaa.edu.cn

$$\frac{dS_t}{S_t} = rdt + \sigma dB_t \qquad (1)$$

where $r$ is risk-free interest rate, $\sigma$ is volatility, $B_t$ is standard Brownian motion and $F_t = \sigma(B_s, 0 \le s \le t)$.

At option expiration time $T$, payoff value of the European call option is $V = (S_T - K)^+$, where $K$ is the strike price, $S_T$ is the assets price at time $T$.

By the no-arbitrage theory, the value of a traditional European call option price is stated as

$$C = SN(d_1) - Ke^{-r(T-t)}N(d_2) \qquad (2)$$

where

$$d_1 = \frac{\ln\frac{S}{K} + (r + \frac{\sigma^2}{2})(T-t)}{\sigma\sqrt{T-t}}, \quad d_2 = \frac{\ln\frac{S}{K} + (r - \frac{\sigma^2}{2})(T-t)}{\sigma\sqrt{T-t}} = d_1 - \sigma\sqrt{T-t}$$

For studying $\sigma$ more convenience, we denote $\tau = T - t$ as the time the option expires, then the formula (2) can be written as

$$C = SN(d_1) - Ke^{-r\tau}N(d_2)$$

where

$$d_1 = \frac{\ln\frac{S}{K} + (r + \frac{\sigma^2}{2})\tau}{\sigma\sqrt{T-t}}, \quad d_2 = d_1 - \sigma\sqrt{\tau}$$

**2.2 European power option pricing formula**

In order to dominate the competition and attract more customers, financial engineers use option theory and analysis methods to design a variety of options with different characteristics of new varieties. According to the needs of the financial market, there are many types of innovative options, power options is a new option type.

Power option is a simple non-linear payment options. We take the power call option for the study, there are two payment forms for $\alpha$-power ($\alpha > 0$) option with option expiration $T$ and strike price $K$

$$V = \left(S_T^{\alpha} - K\right)^+ \qquad (3)$$

$$V = \left(S_T^{\alpha} - K^{\alpha}\right)^+ \qquad (4)$$

we name formula (3) as the first European power call options, and formula (4) as the second European power call options.

The first European call power option pricing formula of formula (3) is as follows [9],

$$C = S^{\alpha} e^{\left[(\alpha-1)r + \frac{1}{2}\alpha(\alpha-1)\sigma^2\right]\tau} N\left(d_1 + (\alpha-1)\sigma\sqrt{t}\right) - Ke^{-rt}N(d_2) \qquad (5)$$

where

$$d_1 = \frac{\ln\left(\frac{S^{\alpha}}{K}\right) + \alpha\left(r + \frac{1}{2}\sigma^2\right)\tau}{\alpha\sigma\sqrt{\tau}}, \quad d_2 = d_1 - \sigma\sqrt{t}$$

And, the second European power call option pricing formula based on formula (4) is as follows [8],

$$C = S^{\alpha} e^{\left[(\alpha-1)r + \frac{1}{2}\alpha(\alpha-1)\sigma^2\right]\tau} N\left(d_1 + (\alpha-1)\sigma\sqrt{t}\right) - K^{\alpha}e^{-rt}N(d_2) \qquad (6)$$

where

$$d_1 = \frac{\ln\left(\frac{S^\alpha}{K^\alpha}\right) + \alpha\left(r + \frac{1}{2}\sigma^2\right)\tau}{\alpha\sigma\sqrt{\tau}}, \quad d_2 = d_1 - \sigma\sqrt{\tau}$$

Obviously, formula (2) takes the special case of formula (5) and formula (6) with $\alpha = 1$.

## 3. Implied volatility formula in European power option
### 3.1 Implied volatility formula in first European power option

For the first European call power options, we use the expansion of the normal distribution function

$$N(x) = \frac{1}{2} + \frac{1}{\sqrt{2\pi}}\left(x - \frac{x^3}{6} + \frac{x^5}{40} + \cdots\right)$$

into the formula(5), and denote $\xi = \sigma\sqrt{\tau}$, $X = Ke^{-r\tau}$. We obtain

$$C = S^\alpha e^{\left[(\alpha-1)r + \frac{1}{2}\alpha(\alpha-1)\sigma^2\right]\tau}\left[\frac{1}{2} + \frac{1}{\sqrt{2\pi}}\left(d_1 + (\alpha-1)\sigma\sqrt{\tau}\right)\right] - X\left[\frac{1}{2} + \frac{1}{\sqrt{2\pi}}\left(d_1 - \sigma\sqrt{\tau}\right)\right]$$

$$= \frac{1}{2}\left(S^\alpha e^{(\alpha-1)r\tau}e^{\frac{1}{2}\alpha(\alpha-1)\xi^2} - X\right) + \frac{1}{\sqrt{2\pi}}\left(S^\alpha e^{(\alpha-1)r\tau}e^{\frac{1}{2}\alpha(\alpha-1)\xi^2} - X\right)\left(\frac{\ln(S^\alpha/K) + \alpha r\tau + \frac{\alpha}{2}\xi^2}{\alpha\xi}\right)$$

$$+ \frac{1}{\sqrt{2\pi}}\left(S^\alpha e^{(\alpha-1)r\tau}e^{\frac{1}{2}\alpha(\alpha-1)\xi^2}(\alpha-1) - X\right)\xi$$

Extending $e^{\frac{1}{2}\alpha(\alpha-1)\xi^2}$ to $e^{\frac{1}{2}\alpha(\alpha-1)\xi^2} = 1 + \frac{1}{2}\alpha(\alpha-1)\xi^2 + \cdots$, we obtain

$$2\sqrt{2\pi}\alpha\xi C =$$

$$S^\alpha e^{(\alpha-1)r\tau}\left(\sqrt{2\pi}\alpha\xi + 2\ln\frac{S^\alpha}{K} + 2r\alpha\tau + \alpha\xi^2 + 2\alpha(\alpha-1)\xi^2 + \alpha(\alpha-1)\ln\frac{S^\alpha}{K}\xi^2 + \alpha(\alpha-1)r\alpha\tau\xi^2\right)$$

$$- X\left(\sqrt{2\pi}\alpha\xi + 2\ln\frac{S^\alpha}{K} + 2r\alpha\tau - \alpha\xi^2\right)$$

Then we can get a quadratic equation about $\xi$

$$\left[S^\alpha\left(\frac{K}{X}\right)^{(\alpha-1)}\left(2\alpha - 1 + \alpha(\alpha-1)(\ln\frac{S}{\sqrt[\alpha]{K}} + r\tau)\right) + X\right]\xi^2 + \sqrt{2\pi}\left[S^\alpha\left(\frac{K}{X}\right)^{(\alpha-1)} - X - 2C\right]\xi$$

$$+2\left[\ln\frac{S}{\sqrt[\alpha]{K}}+r\tau\right]\left(S^\alpha(\frac{K}{X})^{(\alpha-1)}-X\right)=0 \tag{7}$$

Denote $F=S^\alpha(\frac{K}{X})^{(\alpha-1)}$, we get

$$W=2\alpha-1+\alpha(\alpha-1)(\ln\frac{S}{\sqrt[\alpha]{K}}+r\tau)=2\alpha-1+(\alpha-1)\ln\frac{F}{X},$$

Formula (7) changes to

$$(FW+X)\xi^2+\sqrt{2\pi}[(F-X)-2C]\xi+2\left[\frac{1}{\alpha}\ln\frac{F}{X}\right](F-X)=0 \tag{8}$$

Since coefficient $(FW+X)$ could not keep identical sign, the case of largest root becomes very complex.

While $FW+X>0$, as $2\left[\frac{1}{\alpha}\ln\frac{F}{X}\right](F-X)\geq 0$, all real roots of formula (8) are non-negative, the largest root is

$$\xi=\sigma\sqrt{\tau}=\frac{-\sqrt{2\pi}[(F-X)-2C]+\sqrt{\Delta}}{2(FW+X)} \tag{9}$$

only if

$$\begin{cases}\Delta=2\pi[F-X-2C]^2-8[FW+X]\frac{1}{\alpha}\ln\left(\frac{F}{X}\right)(F-X)\geq 0\\ [(F-X)-2C]<0\end{cases}.$$

While $FW+X<0$, as $2\left[\frac{1}{\alpha}\ln\frac{F}{X}\right](F-X)\geq 0$, if the two real roots exist, the largest root is also non-negative. It could be

$$\xi=\sigma\sqrt{\tau}=\frac{-\sqrt{2\pi}[(F-X)-2C]-\sqrt{\Delta}}{2(FW+X)} \tag{10}$$

only if

$$\Delta=2\pi[F-X-2C]^2-8[FW+X]\frac{1}{\alpha}\ln\left(\frac{F}{X}\right)(F-X)\geq 0.$$

**3.2 Implied volatility formula in second European power option**

Similarly, for second European call power options, we use the expansion of the normal distribution function

$$N(x)=\frac{1}{2}+\frac{1}{\sqrt{2\pi}}(x-\frac{x^3}{6}+\frac{x^5}{40}+\cdots)$$

into the formula (6). Denote $\xi = \sigma\sqrt{\tau}$, $X = K^\alpha e^{-r\tau}$, we can get

$$C = S^\alpha e^{\left[(\alpha-1)r+\frac{1}{2}\alpha(\alpha-1)\sigma^2\right]\tau}\left[\frac{1}{2}+\frac{1}{\sqrt{2\pi}}(d_1+(\alpha-1)\sigma\sqrt{\tau})\right] - X\left[\frac{1}{2}+\frac{1}{\sqrt{2\pi}}(d_1-\sigma\sqrt{\tau})\right]$$

$$= \frac{1}{2}\left(S^\alpha e^{(\alpha-1)r\tau}e^{\frac{1}{2}\alpha(\alpha-1)\xi^2} - X\right) + \frac{1}{\sqrt{2\pi}}\left(S^\alpha e^{(\alpha-1)r\tau}e^{\frac{1}{2}\alpha(\alpha-1)\xi^2} - X\right)\left(\frac{\ln(S^\alpha/K^\alpha)+\alpha r\tau + \frac{\alpha}{2}\xi^2}{\alpha\xi}\right)$$

$$+ \frac{1}{\sqrt{2\pi}}\left(S^\alpha e^{(\alpha-1)r\tau}e^{\frac{1}{2}\alpha(\alpha-1)\xi^2}(\alpha-1) + X\right)\xi$$

Expanding the formula $e^{\frac{1}{2}\alpha(\alpha-1)\xi^2}$ with $e^{\frac{1}{2}\alpha(\alpha-1)\xi^2} = 1 + \frac{1}{2}\alpha(\alpha-1)\xi^2 + \cdots$, we can get

$$2\sqrt{2\pi}\alpha\xi C =$$

$$S^\alpha e^{(\alpha-1)r\tau}\left(\sqrt{2\pi}\alpha\xi + 2\ln\frac{S^\alpha}{K^\alpha} + 2r\alpha\tau + \alpha\xi^2 + 2\alpha(\alpha-1)\xi^2 + \alpha(\alpha-1)\ln\frac{S^\alpha}{K^\alpha}\xi^2 + \alpha(\alpha-1)r\alpha\tau\xi^2\right)$$

$$- X\left(\sqrt{2\pi}\alpha\xi + 2\ln\frac{S^\alpha}{K^\alpha} + 2r\alpha\tau - \alpha\xi^2\right)$$

Then, we can also get a quadratic equation about $\xi$,

$$\left[S^\alpha(\frac{K^\alpha}{X})^{(\alpha-1)}\left(2\alpha-1+\alpha(\alpha-1)(\ln\frac{S}{K}+r\tau)\right)+X\right]\xi^2$$

$$+ \sqrt{2\pi}\left[S^\alpha(\frac{K^\alpha}{X})^{(\alpha-1)} - X - 2C\right]\xi + 2\left[\ln\frac{S}{K}+r\tau\right]\left(S^\alpha(\frac{K^\alpha}{X})^{(\alpha-1)} - X\right) = 0 \quad (11)$$

Again, we denote $F = S^\alpha(\frac{K^\alpha}{X})^{(\alpha-1)}$, therefore

$$W = 2\alpha-1+\alpha(\alpha-1)(\ln\frac{S}{K}+r\tau) = 2\alpha-1+(\alpha-1)\ln\frac{F}{X}$$

Then formula (11) changes to

$$(FW+X)\xi^2 + \sqrt{2\pi}[(F-X)-2C]\xi + 2\left[\frac{1}{\alpha}\ln\frac{F}{X}\right](F-X) = 0 \quad (12)$$

Though the variables of $F$ and $X$ are different from those in section 3.1, the formula (12) and formula (8) keep identical form. The same discussion is as follows.

While $FW + X > 0$, as $2\left[\dfrac{1}{\alpha}\ln\dfrac{F}{X}\right](F-X) \geq 0$, all real roots of formula (12) are non-negative, the largest root is

$$\xi = \sigma\sqrt{\tau} = \dfrac{-\sqrt{2\pi}\left[(F-X)-2C\right]+\sqrt{\Delta}}{2(FW+X)} \qquad (13)$$

only if

$$\begin{cases} \Delta = 2\pi\left[F-X-2C\right]^2 - 8[FW+X]\dfrac{1}{\alpha}\ln\left(\dfrac{F}{X}\right)(F-X) \geq 0 \\ \left[(F-X)-2C\right] < 0 \end{cases}.$$

While $FW + X < 0$, as $2\left[\dfrac{1}{\alpha}\ln\dfrac{F}{X}\right](F-X) \geq 0$, if the two real roots exist, the largest root is also non-negative. It could be

$$\xi = \sigma\sqrt{\tau} = \dfrac{-\sqrt{2\pi}\left[(F-X)-2C\right]-\sqrt{\Delta}}{2(FW+X)} \qquad (14)$$

only if $\Delta = 2\pi[F-X-2C]^2 - 8[FW+X]\dfrac{1}{\alpha}\ln\left(\dfrac{F}{X}\right)(F-X) \geq 0$.

Furthermore, the formulae (8)(12) with $\alpha = 1$ will be the corresponding formula of Corrado and Miller's result (1996) in [6].

## 4. Numerical Simulation

Let the original price of the underlying asset $S_0 = 1$ at time $t = 0$, option expiration date $T = 1$, true return standard volatility $\sigma = 15\%$, risk-free interest rate $r = 0.001$, for the strike price, we set $K = 0.9$ (Discount), $K = 1.0$ (Parity), $K = 1.1$ (Premium), and $\alpha \in \{0.4, 0.6, 0.8, 1.0, 1.2, 1.4, 1.6, 1.8, 2.0\}$ respectively. The calculation steps are as follows:

1. The underlying asset price $S_t$ is simulated according to the formula (1), where, the Brownian motion $\Delta B_t \sim N(0, \dfrac{T}{N})$, $N = 100$. For each $\alpha$, the call option price $C(t)$ of two kinds of European option power models are calculated according to the formula (5) and (6) respectively under the realization of $\Delta B_1, \ldots, \Delta B_N$.
2. According to the formulae (9) (10) and formulae (13) (14), we calculate the implied volatility $\hat{\sigma}_i, i = 1, \cdots, N$ at the time $\tau = T - \dfrac{(i-1)}{N}T$; We define three indexes to reflect the random complex of our experiment.

$$dnr = \dfrac{\#\{i \mid \Delta \geq 0, i = 1, 2, \ldots, N\}}{N},$$ which means the existence of roots in formula (8)(12).

$$\bar{\sigma} = \dfrac{1}{L}\sum_{i=1}^{L}\hat{\sigma}_i, \text{ where } L = \#\{i \mid \Delta \geq 0, i = 1, 2, \ldots, N\},$$ which means the average implied volatility for one simulation.

$$\delta_\sigma = \sqrt{\frac{1}{L}\sum_{i=1}^{L}(\hat{\sigma}_i - \bar{\sigma})^2}$$, which measures the divergence degree of volatility estimation in one simulation.

3. Repeat the experiment from step 1 to step 2 for M=100 times, and the average results of $dnr, \bar{\sigma}, \delta_\sigma$ are reported in the Table 1 and Table 2.

**Table 1. Implied volatility estimation of first European call power option**

| K | | $\alpha$ | | | | | | | | |
|---|---|---|---|---|---|---|---|---|---|---|
| | | 0.4 | 0.6 | 0.8 | 1 | 1.2 | 1.4 | 1.6 | 1.8 | 2.0 |
| 0.9 | $dnr$ | 0.0093 | 0.0518 | 0.1335 | 0.2198 | 0.3056 | 0.3646 | 0.4067 | 0.4450 | 0.4742 |
| | $\sigma$ | **0.1346** | 0.1277 | 0.1282 | **0.1298** | 0.1274 | 0.1354 | 0.1394 | 0.1426 | **0.1461** |
| | $\delta_\sigma$ | 0.0163 | 0.0253 | 0.0243 | 0.0231 | 0.0264 | 0.0167 | 0.0130 | 0.0107 | 0.0083 |
| 1.0 | $dnr$ | 0.5163 | 0.5209 | 0.5257 | 0.5314 | 0.5360 | 0.5408 | 0.5476 | 0.5539 | 0.5595 |
| | $\sigma$ | 0.1335 | 0.1357 | 0.1379 | **0.1401** | 0.1425 | 0.1450 | 0.1473 | **0.1498** | 0.1524 |
| | $\delta_\sigma$ | 0.0171 | 0.0150 | 0.0128 | 0.0108 | 0.0085 | 0.0066 | 0.0053 | 0.0048 | 0.0055 |
| 1.01 | $dnr$ | 0.5105 | 0.5248 | 0.5300 | 0.5359 | 0.5414 | 0.5464 | 0.5530 | 0.5584 | 0.5633 |
| | $\sigma$ | 0.1320 | 0.1349 | 0.1373 | **0.1397** | 0.1421 | 0.1446 | 0.1470 | **0.1497** | 0.1524 |
| | $\delta_\sigma$ | 0.0187 | 0.0157 | 0.0134 | 0.0110 | 0.0089 | 0.0067 | 0.0053 | 0.0045 | 0.0053 |

**Table 2. Implied volatility estimation of second European call power option**

| K | | $\alpha$ | | | | | | | | |
|---|---|---|---|---|---|---|---|---|---|---|
| | | 0.4 | 0.6 | 0.8 | 1 | 1.2 | 1.4 | 1.6 | 1.8 | 2.0 |
| 0.9 | $dnr$ | 0.2075 | 0.2113 | 0.2151 | 0.2198 | 0.2246 | 0.2299 | 0.2362 | 0.2425 | 0.2493 |
| | $\sigma$ | 0.1250 | 0.1269 | 0.1290 | **0.1298** | 0.1312 | 0.1332 | 0.1334 | 0.1350 | **0.1359** |
| | $\delta_\sigma$ | 0.0277 | 0.0256 | 0.0233 | 0.0231 | 0.0223 | 0.0201 | 0.0208 | 0.0193 | 0.0192 |
| 1.0 | $dnr$ | 0.5163 | 0.5209 | 0.5257 | 0.5314 | 0.5360 | 0.5408 | 0.5476 | 0.5539 | 0.5595 |
| | $\sigma$ | 0.1335 | 0.1357 | 0.1379 | **0.1401** | 0.1425 | 0.1450 | 0.1473 | **0.1498** | 0.1524 |
| | $\delta_\sigma$ | 0.0171 | 0.0150 | 0.0128 | 0.0108 | 0.0085 | 0.0066 | 0.0053 | 0.0048 | 0.0055 |
| 1.01 | $dnr$ | 0.5178 | 0.5241 | 0.5299 | 0.5359 | 0.5422 | 0.5475 | 0.5545 | 0.5590 | 0.5658 |
| | $\sigma$ | 0.1331 | 0.1352 | 0.1374 | **0.1397** | 0.1420 | 0.1445 | 0.1470 | **0.1498** | 0.1525 |
| | $\delta_\sigma$ | 0.0174 | 0.0154 | 0.0132 | 0.0110 | 0.0089 | 0.0068 | 0.0050 | 0.0042 | 0.0051 |

From the above simulations, we can conclude that the $dnr$ index reflect the successful estimation probability in European call power option pricing model, as the complex of stochastic environment, the $dnr$ with $K=0.9$ is small than that of $K=1$ and $K=1.01$ in both two kind of European call power option. The accuracy estimation of volatility of power option price is slight higher in range of $\alpha>1$ than the case of $\alpha=1$. However, there is still the case that the volatility estimation is more accurate than that with $\alpha=1$, for example, in Table 1, when $\alpha=0.4$, $K=0.9$, though its dnr index is very low. The experimental results partly support the conclusion of [8].

Some further investigation in our research show that if modifying the value of $\Delta$ to guarantee $\Delta\geq 0$ as discussed in [6], the volatility deviation degree will rise. Therefore, we report the $dnr$ index to reflect the effectiveness of formula(8)(12) with power option price. And, the accuracy of volatility estimation inversely reflects the fitting degree with different power option. From Table 1 and Table 2, we can conclude that there exists power option model better than traditional option price in implied volatility estimation. And, the appreciate power index selection will be our further research interest.

## 5. Conclusion

In this paper, with the quadratic Taylor approximations proposed by Corrado and Miller (1996), we derive the close formula of implied volatility in two kind of European call power option pricing, the simulation with Monte-Carlo method also show the effectiveness of our model in implied volatility estimation. The future work will focus on the power option pricing selection and applied our model to real option data application.